\documentclass[showpacs,11pt]{article}
\usepackage{eurosym}
\usepackage{amsmath}
\usepackage{amssymb}
\usepackage{amsmath,amsfonts,latexsym,graphicx,amssymb}
\usepackage{amsfonts}
\usepackage{amssymb}
\usepackage{mathrsfs}
\usepackage{amsthm}
\usepackage{graphicx}
\usepackage{amsmath}
\newcommand{\ot}{{\,\otimes\,}}
\newcommand{{\Cd}}{{\mathbb{C}^d}}
\newcommand{{\Rn}}{{\mathbb{R}^n}}

\def\oper{{\mathchoice{\rm 1\mskip-4mu l}{\rm 1\mskip-4mu l}{\rm 1\mskip-4.5mu l}{\rm 1\mskip-5mu l}}}
\def\<{\langle}
\def\>{\rangle}
\newcommand{\ket}[1]{\left|#1\right\rangle}

\newtheorem{thm}{Theorem}

\newtheorem{ex}{Example}

\newtheorem{pro}{Proposition}
\newtheorem{remark}{Remark}

\begin{document}

\date{}

\title{\textbf{Interaction free and decoherence free states}}
\author{Dariusz Chru\'sci\'nski$^1$, Anna Napoli$^2$, Marina Guccione$^2$, Pawe{\l} Nale\.zyty$^1$ and Antonino Messina$^2$   \\
$^1$Institute of Physics, Faculty of Physics, Astronomy and Informatics, \\
Nicolaus Copernicus University, Grudzi{a}dzka 5, 87--100 Toru\'n, Poland\\
$^2$Dipartimento di Fisica e Chimica
via Archirafi 36, 90123 Palermo, Italy  }
\maketitle

\begin{abstract}
An interaction free evolving state of a closed bipartite system composed of two interacting subsystems is a generally mixed state evolving as if the interaction were a c-number. In this paper we find the characteristic equation of states possessing similar properties for a bipartite systems  governed by a linear dynamical equation whose generator is sum of a free term and an interaction term. In particular in the case of a small system coupled to its environment, we deduce the characteristic equation  of decoherence free states namely mixed states evolving as if the interaction term were effectively inactive. Several examples illustrate the applicability of our theory in different physical contexts.
\end{abstract}

\begin{center}
{  \bf Dedicated to Margarita and Volodia Man'ko for their 150th birthday}
\end{center}

\section{Introduction}

The dynamics of a closed quantum system is fully determined by the system Hamiltonian $H$. Usually, one has a natural splitting $H=H_0+H_I$, with $H_0$ being the free Hamiltonian and $H_I$ the interaction term. If $H_I$ is {\em small} with respect to $H_0$ one treats $H_I$ as a perturbation term and performs well known perturbation expansion. In this paper we analyze a different problem: we look for states $\psi \in \mathcal{H}$ such that 
\begin{equation}\label{I}
  e^{-iHt} \psi = e^{-i\alpha t} e^{-iH_0t} \psi\ ,
\end{equation}
that is, up to a phase factor `$e^{-i\alpha t}$' the total Hamiltonian $H$ and the free part $H_0$ generate the same evolution of $\psi$.  Note, that when $[H_0, H_I]=0$, that is $H$ defines a non-demolition Hamiltonian model, each eigenvector of $H_I$, i.e.  $H_I\psi_\alpha=\alpha\psi_\alpha$, satisfies (\ref{I}). Indeed, one has
\begin{equation}\label{II}
  e^{-iHt} \psi_\alpha = e^{-iH_0t} e^{-iH_It}\psi_\alpha = e^{-i\alpha t} e^{-iH_0t} \psi_\alpha\ . 
\end{equation}
One might wonder whether one can relax commutation of $H_0$ and $H_I$ and still have states satisfying condition (\ref{I}). Recently it has been found \cite{IFE} a positive answer to this question in the bipartite scenario. It was shown \cite{IFE}  that in the Hilbert space of some closed bipartite systems there exists a non-empty subspace $\mathcal{H}_{\rm IFE}$ of $\mathcal{H}$ constituted by  eigenvectors of $H_I$ generally evolving as if $H_I$ were a c-number. The states belonging to $\mathcal{H}_{\rm IFE}$ have been called {\em Interaction-Free Evolving} states (IFE states in short) and when they belong to the kernel of $H_I$, they evolve as if the two subsystems of $S$ were fully decoupled. Generally speaking an IFE state is a non separable mixed state of $S$ unitarily evolving in $\mathcal{H}$ under the action of $H_0$ only and necessary a sufficient condition for the existence of pure or mixed IFE states of a given bipartite closed system have been reported in Ref. \cite{IFE}.

In this paper we generalize the concept of IFE states problem to arbitrary  ``systems" whose dynamics is characterized by a linear first order homogeneous differential equation whose generator $L$ allows to a natural splitting $L=A+B$. We ask for the existence of states which are unaffected by $B$, i.e. their evolution is fully determined by $A$. In particular we explore the extension of the definition of IFE state to open quantum systems whose evolution is governed by the Markovian master equation. In this case $B$-part of $L$ corresponds to the dissipative/decoherence part of the generator. States which are unaffected by $B$ evolve in a unitary way without any decoherence effects. Due to such property, "IFE" states in this context will be called {\em decoherence-free states}.
It is clear that such states are closely related to well known concept of decoherence free subspaces \cite{zanardi1,zanardi2,Lidar,lidar2013, Schlosshauer, Nalezyty} and subradiant states \cite{subradiant1, subradiant2, subradiant3, subradiant4, subradiant5}. Our main result is the determination of characteristic equation of such mixed states for an open system. We illustrate the usefulness of our mathematical conditions explicitly constructing examples of both IFE and decoherence free states in various physical scenarios.

The paper is organized as follows: in the next section we provide the general scheme for evolution governed by the linear generator $L=A+B$ and define states which are interaction free or more precisely $B$-free. Section \ref{INT} analyzes the structure of IFE states in the interaction picture -- it turns out that interaction picture evolution of IFE states is trivial. Section \ref{EX} shows how the Schr\"odinger evolution fits the general scheme. This section is illustrated by several examples. Section \ref{OPEN} generalizes Schr\"odinger evolution of a closed system to Markovian evolution of open quantum systems and introduces the concept of decoherence free states. Final conclusions are collected in section \ref{CON}.

\section{General scheme} \label{GEN}

Consider a linear dynamical equation
\begin{equation}\label{D}
  \dot{\mathbf{x}}_t = (A + B)\mathbf{x}_t\ ,\ \ \ \mathbf{x}_{t=0} = \mathbf{x}\ ,
\end{equation}
where $\mathbf{x}_t$ is a vector in a linear space $W$ and $A,B : W \rightarrow W$ are linear operators.
$W$ might be interpreted as a space of states of some physical system and $L= A+B$ plays the role of the generator which is divided into two parts: $A$ -- usually referred as a ``free" part and $B$ -- referred as an ``interaction" part. Typically, in physical applications $W$ is a Hilbert space $\mathcal{H}$ or an algebra of bounded operators $\mathcal{B}(\mathcal{H})$ or a Banach space of trace class operators $\mathcal{T}(\mathcal{H})$.  The formal solution to (\ref{D}) is given by
 $ \mathbf{x}_t = e^{(A+B)t} \mathbf{x}$.
We call $\mathbf{x}$ a $B$-free state if
\begin{equation}\label{B-free}
e^{(A+B)t} \mathbf{x} = e^{At} \mathbf{x}
\end{equation}
for all $t$.

\begin{thm}\label{TH} A state $\mathbf{x}$ is $B$-free if and only if
\begin{equation}\label{}
  \mathbf{x} \in \mathcal{M} := {\rm Ker}\, B \cap  {\rm Ker}\, BA \cap \ldots \cap {\rm Ker}\, BA^{n-1}\ ,
\end{equation}
where $n = {\rm dim}\,W$.
\end{thm}
Proof: expanding $e^{t(A+B)}$ and $e^{tA}$ one finds that $\mathbf{x}$ is $B$-free iff
\begin{equation}\label{==}
  \left(\mathbb{I} + t (A+B) + \frac{t^2}{2}(A^2 + AB +BA + B^2) + \ldots \right) \mathbf{x} =
  \left(\mathbb{I} + t A + \frac{t^2}{2}A^2 + \ldots \right) \mathbf{x}\ ,
\end{equation}
for all $t$. It is clear that if $\mathbf{x} \in \mathcal{M}$ then (\ref{==}) holds. Conversely, if condition (\ref{==}) is satisfied, then differentiating both sides at $t=0$ one hinds
$$  (A+B)\mathbf{x} = A\mathbf{x}\ , $$
which shows that $\mathbf{x} \in {\rm Ker}\, B$. Performing the 2nd derivative at $t=0$ gives
$$ (A^2+AB+BA+B^2)\mathbf{x} = A^2\mathbf{x}\ , $$
and taking into account that $\mathbf{x} \in {\rm Ker}\, B$ one finds that $\mathbf{x} \in {\rm Ker}\, BA$. Continuing this process up to the $n$th derivative one finally recovers the definition of $\mathcal{M}$. \hfill $\Box$

\begin{remark} It is clear that if $[A,B]=0$, then $e^{(A+B)t} = e^{At} e^{Bt}$ and  $ \mathcal{M} = {\rm Ker}\, B$.
If $[A,B] \neq 0$, then $\mathcal{M}$ is a proper subspace of ${\rm Ker}\, B$. Note, that $\mathcal{M}$ is A,B-invariant,
i.e. $A\mathcal{M} \subset \mathcal{M}$ and  $B\mathcal{M} \subset \mathcal{M}$. One has
\begin{equation}\label{}
  [A,B]\Big|_\mathcal{M} = 0 \ ,
\end{equation}
i.e. $A$ and $B$ are partially commuting. In this case one has
\begin{equation}\label{}
  e^{(A+B)t}\Big|_\mathcal{M} = e^{At} e^{Bt}\Big|_\mathcal{M}\ .
\end{equation}

\end{remark}

\begin{remark} Interestingly, it was shown by Shemesh \cite{Shemesh} that $\mathcal{M} \neq 0$ if and only if there exists $\mathbf{x}_0$
such that
\begin{equation}\label{AB0}
  A\mathbf{x}_0 = a\mathbf{x}_0\ , \ \ \ B\mathbf{x}_0=0 \ .
\end{equation}
Clearly, $\mathbf{x}_0$
satisfying (\ref{AB0}) is necessarily $B$-free
\begin{equation}\label{}
  e^{(A+B)t} \mathbf{x}_0 = e^{At} \mathbf{x}_0 = e^{at} \mathbf{x}_0\ .
\end{equation}
Any $B$-free
state $\mathbf{x}$ satisfies $B\mathbf{x}=0$. However, in general it needs not be eigenvector of $A$.
\end{remark}

\begin{remark}
Usually, one interprets $A$ as the ``free'' generator and $B$ as the ``interaction'' part. In this case $B$-free states may be called {\em interaction-free} (IFE) \cite{IFE}. Actually, in \cite{IFE} the following situation was considered: $W = \mathcal{H}_1 \ot \mathcal{H}_2$ together with $A = H_1 \ot \mathbb{I}_2 + \mathbb{I}_1 \ot H_2$ and $B= H_I$ denotes the interaction Hamiltonian. The evolution of the interaction free state does not depend upon the interaction part $H_I$. The characteristic feature of IFE states is the conservation of energy of two subsystems
\begin{equation}\label{}
  \mathcal{E}_1(t) = \< \psi_{12}(t) |H_1 \ot \mathbb{I}_2| \psi_{12}(t)\>\ , \ \ \   \mathcal{E}_2(t) = \< \psi_{12}(t) | \mathbb{I}_1 \ot H_2| \psi_{12}(t)\> \ ,
\end{equation}
that is, $\mathcal{E}_1(t)$ and $\mathcal{E}_2(t)$ are time-independent.
\end{remark}

\section{Interaction free states -- interaction picture}   \label{INT}

 Passing to the ``interaction picture"
\begin{equation}\label{}
  {\mathbf{y}}_t = e^{-At} \mathbf{x}_t\ ,
\end{equation}
one finds
\begin{equation}\label{}
  \dot{{\mathbf{y}}}_t  = {B}(t) \mathbf{y}_t\ , \ \ \ \mathbf{y}_{t=0} =\mathbf{x}\ ,
\end{equation}
with the time-dependent $B$-part
\begin{equation}\label{}
  B(t) = e^{-At} B e^{At}\ .
\end{equation}
The formal solution reads
\begin{equation}\label{}
  \mathbf{y}_t = T \exp\left( \int_0^t{B}(\tau)d\tau \right)\mathbf{x} = \left(  \oper + \int_0^t dt_1 {B}(t_1) +
  \int_0^t dt_1 \int_0^{t_1} dt_2 {B}(t_1){B}(t_2) + \ldots \right) \mathbf{x}\ .
\end{equation}

\begin{pro}
A state $\mathbf{x}$ is $B$-free if
\begin{equation}\label{}
  \mathbf{y}_t = \mathbf{x}\ ,
\end{equation}
that is, the interaction picture solution $\mathbf{y}_t$ is trivial.
\end{pro}
Proof:  using Baker-Campbell-Hausdorff formula
\begin{equation}\label{}
  e^{-At} B e^{At} = B + t [B,A] + \frac{t^2}{2} [[B,A],A] +  \frac{t^3}{3!} [[[B,A],A],A] + \ldots
\end{equation}
one easily checks that
\begin{equation}\label{}
  B(t)\Big|_\mathcal{M} =  e^{-At} B e^{At} \Big|_\mathcal{M} = 0 \ ,
\end{equation}
and hence states from $\mathcal{M}$ belong to the kernel of $B(t)$. \hfill $\Box$

Interaction picture may be also called a ``$B$-picture" (in $B$-picture the $A$ part is eliminated). Equivalently, we may
introduce an ``$A$-picture" by eliminating $B$ part: introducing
\begin{equation}\label{}
  {\mathbf{z}}_t = e^{-Bt} \mathbf{x}_t\ ,
\end{equation}
one finds
\begin{equation}\label{}
  \dot{{\mathbf{z}}}_t  = {A}(t) \mathbf{z}_t\ , \ \ \ \mathbf{z}_{t=0} =\mathbf{x}\ ,
\end{equation}
with time-dependent $A$-part
\begin{equation}\label{}
  A(t) = e^{-Bt} A e^{Bt}\ .
\end{equation}

\begin{pro}
A state  $\mathbf{x}$ is $B$-free if
\begin{equation}\label{}
  \mathbf{z}_t = \mathbf{x}_t\ ,
\end{equation}
that is, $A$-picture solution $\mathbf{z}_t$ coincides with the original solution $\mathbf{x}_t$.
\end{pro}
Again, BCH formula implies
\begin{equation}\label{}
  A(t)\Big|_\mathcal{M} = A\Big|_\mathcal{M}\ ,
\end{equation}
and the result immediately follows.

\section{Schr\"odinger evolution}   \label{EX}

Consider now the Schr\"odinger equation
\begin{equation}\label{}
  i\dot{\psi}_t = (H_0 + H_I)\psi_t\ , \ \ \ \psi_{t=0} = \psi\ .
\end{equation}
Since any two normalized vectors $\psi$ and $\phi$ such that $\phi = e^{i\alpha} \psi$ define the same physical state  we call $\psi$
an interaction free state if
\begin{equation}\label{}
  e^{-i(H_0+H_I)t}\psi = e^{-i\alpha t} e^{-iH_0t}\psi\ ,
\end{equation}
for some real $\alpha$. Let $H_I^{(\alpha)} = H_I - \alpha \oper$.

\begin{thm}[\cite{IFE}] A vector state $\psi$ is interaction free if and only if
\begin{equation}\label{}
  \psi \in \mathcal{M}_\alpha := {\rm Ker}\,H_I^{(\alpha)}  \cap  {\rm Ker}\, H_I^{(\alpha)}H_0 \cap \ldots \cap {\rm Ker}\,
  H_I^{(\alpha)}H_0^{n-1}\ ,
\end{equation}
where $n = {\rm dim}\,\mathcal{H}$.
\end{thm}
Clearly $\alpha$ defines an eigenvalue of $H_I$. Note that $\mathcal{M}_\alpha $ is non-trivial iff there exists common eigenvector
of $H_0$ and $H_I$
$$    H_0 \psi_0 = \lambda \psi_0\ , \ \ \ H_I \psi_0 = \alpha\psi_0\ . $$

Shemesh theorem \cite{Shemesh} states that  $H_0$ and $H_I$  have a common eigenvector if and only if
\begin{equation}\label{}
  \mathcal{M} = \bigcap_{k,l=1}^{n-1} {\rm Ker}[H_0^k,H_I^l]\
\end{equation}
is a nontrivial subspace of $\mathcal{H}$. If $\{\alpha_1,\alpha_2,\ldots,\alpha_n\}$ defines the spectrum of $H_I$, then
\begin{equation}\label{}
  \mathcal{M} = \mathcal{M}_{\alpha_1} \oplus \ldots \oplus \mathcal{M}_{\alpha_n} \ .
\end{equation}
The subspace $\mathcal{M}$ is a maximal common invariant subspace of $H_0$ and $H_I$. Taking
$$ \psi = \psi_1 \oplus \ldots \oplus \psi_n \ , $$
with $\psi_k \in \mathcal{M}_{\alpha_k}$ one finds
\begin{equation}\label{}
  \psi_t = e^{-iH_0t} [ e^{-i\alpha_1 t} \psi_1 \oplus \ldots \oplus e^{-i\alpha_n t} \psi_n ] \ .
\end{equation}
Note, that $\psi_t$  belongs to $\mathcal{M}$ but $\psi$ is not interaction free unless $\psi$ belongs to a single ``sector" $ \mathcal{M}_{\alpha_k}$.

\begin{remark} Since $H_0$ and $H_I$ are Hermitian and commute on $\mathcal{M}$ there exists an orthonormal basis $\{|1\>,\ldots,|m\>\}$ on $\mathcal{M}$  such that
\begin{equation}\label{}
  H_0\Big|_{\mathcal{M}} = \sum_{k=1}^m a_k |k\>\<k| \ , \ \ \   H_I\Big|_{\mathcal{M}} = \sum_{k=1}^m b_k |k\>\<k|\ .
\end{equation}
Note, that it is no longer true in the general case with arbitrary operators $A$ and $B$ unless they are normal. In the general case $A\Big|_{\mathcal{M}}$ and $B\Big|_{\mathcal{M}}$ have the same Jordan block structure.
\end{remark}

Schr\"odinger evolution may be generalized for mixed states represented by density operators. Consider the von Neumann equation
\begin{equation}\label{}
  \dot{\rho}_t = (L_0 + L_I)\rho_t\ ,
\end{equation}
where $L_0 \rho = - i[H_0,\rho]$ and  $L_I \rho = - i[H_I,\rho]$.

\begin{pro} A mixed state $\rho$ is interaction free if and only if
\begin{equation}\label{}
  \rho \in \mathcal{M} := {\rm Ker}\,L_I  \cap  {\rm Ker}\, L_I L_0 \cap \ldots \cap {\rm Ker}\,
  L_I L_0^{n^2-1}\ ,
\end{equation}
and $\mathcal{M} \cap S(\mathcal{H}) \neq 0$, where ${S}(\mathcal{H}) = \{ \rho \, |\, \rho \geq 0\, , \ {\rm tr}\rho=1\}$.
\end{pro}
Note, that any interaction free state may be represented as follows
\begin{equation}\label{}
  \rho = \rho_1 \oplus \ldots \oplus \rho_n \ ,
\end{equation}
where $\rho_k$ is supported on $ \mathcal{M}_{\alpha_k}$. Indeed, one finds
\begin{equation}\label{}
  \rho_t = e^{-i H_0 t}\Big|_{\mathcal{M}_{\alpha_1}} \rho_1\,  e^{H_0 t}\Big|_{\mathcal{M}_{\alpha_1}}
   \oplus \ldots \oplus \, e^{-i H_0 t}\Big|_{\mathcal{M}_{\alpha_n}} \rho_n\,  e^{H_0 t}\Big|_{\mathcal{M}_{\alpha_n}}\ ,
\end{equation}
that is, $\rho$ is supported on $\mathcal{M}$ has block-diagonal structure. Diagonal blocks $\rho_k$ are supported on
$ \mathcal{M}_{\alpha_k}$ and each diagonal block evolves independently. One finds the similar block structure for $H_0\Big|_{\mathcal{M}}$
\begin{equation}\label{}
  H_0\Big|_{\mathcal{M}} = h_1 \oplus \ldots \oplus  h_n\ ,
\end{equation}
where
\begin{equation}\label{}
  h_k := H_0\Big|_{\mathcal{M}_{\alpha_k}}\ .
\end{equation}
Hence
\begin{equation}\label{}
  \rho_t = e^{-i h_1 t} \rho_1\,  e^{i h_1 t}
   \oplus \ldots \oplus \, e^{-i h_n t} \rho_n\,  e^{i h_n t}\ .
\end{equation}
Note, that if $\rho$ is pure, then $\rho_k = \delta_{km} P$ for some $m$ and $P=|\psi\>\<\psi|$. In this case one has $\rho_t = e^{-i h_m t} \rho_m\,  e^{i h_m t}$.

\begin{ex}    Let us consider a toy model, traceable back to the so-called intensity dependent Jaynes-Cummings models \cite{BS, BMN}, composed by a  single bosonic mode coupled to a two level atom as described by the following Hamiltonian
\begin{equation}\label{1}
 H=\omega[a^\dag a+ (-1)^N\frac{1}{2}\sigma_z]=H_0+H_I
\end{equation}
where $H_0=\omega a^\dag a$, $H_I=\omega (-1)^N \frac{1}{2}\sigma_z$ and $N=a^\dag a+ \frac{1}{2}(1+\sigma_z)$. One has
\begin{equation}\label{2}
[H_0,H_I]=0\ ,
\end{equation}
and hence $\mathcal{M}$ is a direct sum of eigenspaces of $H_I$. Moreover, the number operator $N$ commutes with the interaction Hamiltonian $[N,H_I]=0$. Note, that states $\ket{n, \sigma} $ defined as
$$ a^\dag a \ket{n, \sigma}=n\ket{n, \sigma}\ , \ \ \ \sigma_z \ket{n, \sigma}=\sigma\ket{n, \sigma}\ , \  \ \ \sigma = \pm 1\ ,$$
are common eigenstates of $H_0$ and $H_I$. One has
$$   N|n,+\> = (n+1)|n,+\>\ , \ \ \ N|n,-\>=n|n,-\>\ , $$
and hence
$$  (-1)^N|n,+\> = (-1)^{n+1}|n,+\>\ , \  \ (-1)^N|n,-\> = (-1)^{n+1}|n,-\>\ . $$
Finally
\begin{equation}\label{4}
H_I\ket{n,+}= \frac{\omega}{2}(-1)^{n+1} \ket{n,+} \;\;\;\;\:\: H_I\ket{n,-}=\frac{\omega}{2} (-1)^n \ket{n,-}\ ,
\end{equation}
which proves that $H_I$ has only two eigenvalues $\alpha_\pm = \pm \frac{\omega}{2}$, that is,
\begin{eqnarray*}
  \mathcal{N}_{\alpha_+} &=& {\rm span} \{ \ |2n+1,-\>, |2n+1,+\> \ ;  \ n=0,1,2,\ldots \  \}\ , \\
  \mathcal{N}_{\alpha_-} &=& {\rm span} \{ \ |2n,-\>, |2n,+\> \ ;  \ n=0,1,2,\ldots \  \}\ .
\end{eqnarray*}
Obviously, they are eigenspaces of the parity operator $(-1)^N$:
$$   (-1)^N\mathcal{N}_{\alpha_\pm} = \pm \mathcal{N}_{\alpha_\pm} \ . $$
Clearly, both eigenvalues are infinitely degenerated:
\begin{eqnarray*}
  \mathcal{N}_{\alpha_+} = \mathcal{N}_{+}^0 \oplus \mathcal{N}_{+}^1 \oplus \ldots\ ,   \ \
  \mathcal{N}_{\alpha_-} = \mathcal{N}_{-}^0 \oplus \mathcal{N}_{-}^1 \oplus \ldots \ ,
\end{eqnarray*}
where $\mathcal{N}_{+}^n \simeq \mathbb{C}^2$ and $ \mathcal{N}_{-}^n \simeq \mathbb{C}^2$ are defined by
\begin{eqnarray*}
  \mathcal{N}_{+}^n = {\rm span} \{ \ |2n+1,-\>, |2n+1,+\>  \  \}\ , \ \   \mathcal{N}_{-}^n = {\rm span} \{ \ |2n,-\>, |2n,+\>  \  \}\ .
\end{eqnarray*}
Consider for example $|\psi_+\> \in \mathcal{N}_{\alpha_+}$, that is,
\begin{equation}\label{6}
\ket{\psi_+}=  \sum_n [ a_n \ket{2n+1,-} + b_n \ket{2n+1,+} ]
\end{equation}
with $\sum_n [ |a_n|^2 + |b_n|^2]=1$. In this case one finds for the evolution
 \begin{equation}\label{7}
e^{-iHt}\ket{\psi}=  e^{-i\frac{\omega}{2}t} e^{-iH_0t} \ket{\psi} = e^{-i\frac{\omega}{2}t} \sum_n e^{-i\omega(2n+1)t} [ a_n  \ket{2n+1,-} + b_n  \ket{2n+1,+} ]\ .
\end{equation}
Clearly, $e^{-iHt}\ket{\psi}$ defines a nontrivial trajectory in $\mathcal{N}_{\alpha_+}$.
\end{ex}

\begin{ex} Let us consider the two-mode Jaynes-Cummings model \cite{JC,OP, BM} described by the following Hamiltonian
\begin{equation}\label{8}
 H=\omega(a_1^\dag a_1+a_2^\dag a_2)+ \omega_0 \frac{1}{2}\sigma_z+ \sum_{i=1,2}\gamma_i (a_i^\dag \sigma_-+a_i \sigma_+)=H_0+H_I
\end{equation}
For simplicity consider the resonant case, i.e. $\omega_0=\omega$. One easily checks that  the free Hamiltonian $H_0$ and the interaction Hamiltonian $H_I$ commute. Moreover the number operator $N=a_1^\dag a_1+a_2^\dag a_2+ \frac{1}{2}(1+\sigma_z)$ is a constant of motion and hence $[H_I, N]=0$. One finds
\begin{equation} \label{N_eigenstates}
   N|n_1,n_2,+\> = (n_1+n_2 +1)|n_1,n_2,+\>\ , \; \; \; \; \; N|n_1,n_2,-\> = (n_1+n_2)|n_1,n_2,-\>\
\end{equation}
Starting from eq. (\ref{N_eigenstates}) it is easy to demonstrate that the eigenspace $\mathcal{H}_n$ corresponding to the  eigenvalue $n$ of $N$ is $(2n+1)$-dimensional. Let's indeed observe that the set $S=S_-\cup S_+$ where
\begin{equation}\label{s-}
S_-=\{\ket{n_1,n_2,-},\;\; n_i=0,..,n \;\;\;(i=1,2) \;\;\;\; with \;\;\; n_1+n_2=n\}
\end{equation}
contains $n+1$ orthogonal states and
\begin{equation}\label{s+}
S_+= \{ \ket{m_1,m_2,+}, \;\; m_i=0,..,n-1 \;\;\;(i=1,2)\;\;\;\;  with\;\;\;  m_1+m_2+1=n\}
\end{equation}
which contains instead  $n$ states,  forms a basis of $\mathcal{H}_n$. We may diagonalize the interaction Hamiltonian $H_I$ in any $(2n+1)$-dimensional invariant subspace of $N$.

Let us consider in particular the two Hilbert subspaces correspondent to $n=1$ and $n=2$ respectively. Diagonalizing $H_I$ in the subspace with $n=1$ we obtain three distinct eigenvalues, namely  $(0, \;  -\sqrt{\gamma_1^2+\gamma_2^2},\;  \sqrt{\gamma_1^2+\gamma_2^2})$. In the subspace with $n=2$ we instead  obtain  the following eigenvalues

$$ \left\{0, \;  -\sqrt{\gamma_1^2+\gamma_2^2},\;  \sqrt{\gamma_1^2+\gamma_2^2},  -\sqrt{2}\sqrt{\gamma_1^2+\gamma_2^2},\;  \sqrt{2}\sqrt{\gamma_1^2+\gamma_2^2} \right\}\ . $$
Thus, if we consider the eigenvalue of $H_I$  given by $\alpha=\sqrt{\gamma_1^2+\gamma_2^2}$ we may easily construct the correspondent eigenvectors and thus we may explicitly give IFE states belonging to  $\mathcal{N}_\alpha$ that are not stationary states.

\end{ex}

\section{Markovian semigroup and decoherence free states}  \label{OPEN}

In this section we generalize the IFE states defined for  unitary evolution to the evolution corresponding to quantum Markovian semigroup \cite{GKS,L,Alicki,Breuer,EPL} (see also recent review \cite{OSID-2014}). A general quantum evolution of a system living in the Hilbert space $\mathcal{H}$ is described by a dynamical map, that is, a family of completely positive trace-preserving maps (CPTP) $\Lambda_t : \mathfrak{B}(\mathcal{H}) \rightarrow \mathfrak{B}(\mathcal{H})$ such that $\Lambda_0 = \oper$ (an identity map). Let us recall that any dynamical map $\Lambda_t$ may be realised as follows: one considers a composed system living in $\mathcal{H}_{\rm total} := \mathcal{H}\otimes \mathcal{H}_E$ and the unitary evolution in $\mathcal{H}_{\rm total}$ generated by the total Hamiltonian $H_{\rm total}$. Starting with the initial product state $\rho \ot \omega_E$ one defines a map
\begin{equation}\label{}
  \rho \rightarrow \Lambda_t\rho := {\rm Tr}_E [\,U_t \rho \otimes \omega_E U_t^\dagger\,]\ ,
\end{equation}
where ${\rm Tr}_E$ denotes the partial trace over the environmental degrees of freedom and $U_t = \exp(-i t H_{\rm total})$. By construction $\Lambda_t$ is CPTP \cite{Alicki}. Let us observe that the reduced dynamics (after partial trace) is never unitary which shows that the system is open. Markovian semigroup is defined by the dynamical map $\Lambda_t$ satisfying Markovian master equation
\begin{equation}\label{ME}
  \dot{\Lambda}_t = L \Lambda_t\ , \ \ \Lambda_0 = \oper\ ,
\end{equation}
where $L$ denotes the corresponding generator. The formal solution is given by $\Lambda_t = e^{tL}$ and the semigroup property 
\begin{equation}\label{}
  \Lambda_t \circ \Lambda_\tau = \Lambda_{t+\tau} \ , 
\end{equation}
for arbitrary $t,\tau \geq 0$ immediately follows. It was proved by Gorini et al and Lindblad \cite{GKS,L} that the solution to (\ref{ME}) provides legitimate dynamical map iff the generator has the following form
\begin{equation}\label{}
  L \rho = -i[H_0,\rho] + \frac 12 \sum_k \left( [V_k, \rho V_k^\dagger] + [V_k \rho, V_k^\dagger] \right)\ ,
\end{equation}
where $H_0$ denotes effective system Hamiltonian and $V_k$ arbitrary noise operators. One has a natural splitting $L=L_0 + L_D$, where $L_0\rho = -i[H_0,\rho]$ and $L_D$ stands for the dissipative (or decoherence) part. Dissipative part is responsible for all dissipative/decoherence phenomena which can not be described by the unitary evolution. 

We call $\rho$   {\em decoherence free state} if
\begin{equation}\label{}
  e^{(L_0 + L_D)t} \rho = e^{L_0t} \rho =  e^{-i H_0 t} \rho  e^{H_0 t} \ ,
\end{equation}
for all $t \geq 0$. Decoherence free states evolve in a unitary way.

\begin{pro} A mixed state $\rho$ is decoherence free if and only if
\begin{equation}\label{}
  \rho \in \mathcal{M} = {\rm Ker}\,L_D \cap  {\rm Ker}\, L_D L_0 \cap \ldots \cap {\rm Ker}\,
  L_D L_0^{n^2-1} \ ,
\end{equation}
and  where ${S}(\mathcal{H}) = \{ \rho \, |\, \rho \geq 0\, , \ {\rm tr}\rho=1\}$.
\end{pro}

\begin{ex} Consider pure decoherence of a qubit governed by
\begin{equation}\label{}
  H_0 =  \omega \sigma_z\ , \ \ L_D \rho = \gamma (\sigma_z \rho \sigma_z - \rho)\ .
\end{equation}
Note, that $[L_0,L_D]=0$ and hence hence $\mathcal{M} = {\rm Ker}\, L_D$.
Denoting $P_k = |k\>\<k|$, where $\sigma_z|0\>=|0\>$ and $\sigma_z|1\>=-|1\>$, one finds
\begin{equation}\label{}
  \mathcal{M} \cap S(\mathcal{H})=  \{\, \rho = p_0 P_0 + p_1 P_1 \,|\, p_0+p_1=1\, \}\ ,
\end{equation}
that is, any state being a convex combination of $P_k$ is decoherence free. Note, that taking $H_0 = \omega \sigma_x$ one finds that $L_0$ and $L_D$ no longer commute and $\mathcal{M} = 0$, that is, there is no decoherence free states.
\end{ex}

\begin{ex} Pure decoherence model for a qubit may be generalized for an arbitrary qudit system: let $\{|0\>,\ldots|d-1\>\}$ be an orthonormal basis in $d$-dimensional Hilbert space $\mathcal{H}$. Define a class of unitary Weyl operators
\begin{equation}\label{}
  U_{nm}|k\> = \omega^{nk}|m+k\> \ , \ \ \ {\rm mod}\ d\ ,
\end{equation}
where $\omega = e^{2\pi i/d}$. These operators satisfy
\begin{eqnarray*}
  U_{nm} U_{rs}  &=& \omega^{ms} U_{n+r,n+s}\ , \\
  U_{mn}^\dagger &=& \omega^{mn} U_{-m,-n} \ ,
\end{eqnarray*}
and the following orthogonality relation
\begin{equation*}
  {\rm tr}[ U_{mn}^\dagger U_{kl}] = d \delta_{mk}\delta_{nl}\ .
\end{equation*}
Note, that $U_{00} = \mathbb{I}$ and $U_{n0}$ are diagonal and hence mutually commuting. Define
\begin{equation}\label{}
  H_0 = \sum_{n=1}^{d-1} E_n P_n \ , \ \ L_D\rho = \sum_{n=1}^{d-1} \gamma_n [ U_{n0} \rho U_{n0}^\dagger - \rho] \ ,
\end{equation}
with $\gamma_n \geq 0$, real energy levels $E_n$,  and $P_n = |n\>\<n|$. Note, that again $[L_0,L_D]=0$ and hence $\mathcal{M} = {\rm Ker}\, L_D$.
One finds
\begin{equation*}
  L_D |k\>\<l| = \left\{\sum_{n=0}^{d-1} \gamma_n(\omega^{n(k-l)} - 1)\right\} |k\>\<l|\ ,
\end{equation*}
and $L_0 |k\>\<l| = -i(E_k-E_l)|k\>\<l|$. Hence, $L_0|k\>\<k| = L_D|k\>\<k|=0$. The evolution of the initial density matrix $\rho$ has the following form
\begin{equation}\label{}
  \rho(t) = D(t) \circ \rho\ ,
\end{equation}
where $D\circ \rho$ denotes the Hadamard product and $D_{kl}(t) = \exp\left[ \sum_{n=0}^{d-1} \gamma_n(\omega^{n(k-l)} - 1)\right]$. Note, that $D_{kk}(t) = 1$, and $|D_{kl}(t)| < 1$ for $k\neq l$ due to ${\rm Re}\,\omega^{n(k-l)} < 1$. Therefore, this model describes pure decoherence of a qudit. Finally, one finds
\begin{equation}\label{}
  \mathcal{M} \cap S(\mathcal{H})=  \{\, \rho = p_0 P_0 + \ldots + p_{d-1} P_{d-1} \,|\, p_0+ \ldots + p_{d-1}=1\, \}\ ,
\end{equation}
that is, any state being a convex combination of $P_k$ is decoherence free.
\end{ex}

\begin{ex}[Phase damping of harmonic oscillator]

Let
\begin{equation}\label{}
  L \rho =  -i[H_0,\rho]+\Gamma \left(a^\dagger a \rho {a}^\dagger a - \frac 12 (a^\dagger a)^2 \rho - \frac 12 \rho (a^\dagger a)^2 \right)\ ,
\end{equation}
with $H_0=\omega a^\dagger a$. Again $[L_0,L_D]=0$. Note, that $LN=0$ and hence $N=a^\dagger a$ is a constant of motion. In the energy eigenbasis $N|m\> = m|m\>$ one has
\begin{equation}\label{}
  \rho = \sum_{n,m=0}^\infty \rho_{nm} |n\>\<m|\ ,
\end{equation}
and hence the dynamical map $\Lambda_t = e^{tL}$ reads $\Lambda_t \rho = \sum_{n,m} \rho_{nm} \Lambda_t|n\>\<m|$. One finds
\begin{equation}\label{}
  L_D |n\>\<m| = [-i\omega(n-m) - \frac 12 (n-m)^2 \Gamma] |n\>\<m|\ ,
\end{equation}
and hence the off-diagonal elements are transformed according to
\begin{equation}\label{}
  \rho_{nm} \ \rightarrow\ \exp\{[-i\omega(n-m) - \frac 12 (n-m)^2 \Gamma]t\} \rho_{nm} \ ,
\end{equation}
and asymptotically the density matrix is perfectly decohered in the energy eigenbasis. It is clear that states
\begin{equation}\label{}
  \rho = \sum_{n=0}^\infty \rho_{nn} |n\>\<n|\ ,
\end{equation}
are decoherence free.
\end{ex}


\section{Conclusions}  \label{CON}

In this paper we analyzed a class of linear dynamical equations with the generator $L$ being a sum of two terms $L=A+B$ (a `free' part $A$ and `interaction' (or `dissipation/decoherence' part $B$).  We characterized a class of  states which are insensitive to the B part. In the context of Schr\"odinger evolution of bipartite systems these states were already analyzed in \cite{IFE} as Interaction Free States (IFE). We stress that this problem is very general and may be studied whenever the evolution of the system is linear and its generator has a natural splitting into two parts $A+B$. Interestingly, the necessary and sufficient condition for the existence of such states (cf. Theorem \ref{TH}) is closely related to the existence of common eigenvectors of $A$ and $B$ \cite{Shemesh}. We analyzed Schr\"odinger evolution of pure and mixed states governed by the Hamiltonian $H=H_0 + H_I$ and dissipative evolution of density operator governed by the Markovian generator $L = L_0 + L_D$, with $L_0\rho = -i[H_0,\rho]$ and $L_D$ being a purely dissipative/decoherence part. In the latter case we call state insensitive to $L_D$ -- decoherence free. It is clear that decoherence free free states have to be related to the well known concept of decoherence free subspaces \cite{Lidar}. This issue needs further investigation. It would be also interesting to generalize IFE states for the time-dependent case, i.e. either time-dependent Hamiltonian or time-dependent generator of open system evolution.

\section*{Acknowledgements}

We thank Andrzej Jamio{\l}kowski for pointing out \cite{Shemesh}.
DC was partially supported by the National Science Center project
DEC-2011/03/B/ST2/00136.

\end{document}